# Aspects and challenges of mashup creator design

Lampros Goussis
Department of
Telecommunications Science and
Technology
University of Peloponnese
Tripoli, Greece
gl7@uop.gr

Ioannis E. Foukarakis
Department of
Telecommunications Science and
Technology
University of Peloponnese
Tripoli, Greece
ifouk@uop.gr

Dimitios N. Kallergis
Department of Electronic
Computer Systems Engineering
Technological Educational Institute
(TEI) of Piraeus
Piraeus, Greece
D.Kallergis@teipir.gr

*Abstract*— **With the advent of Web 2.0, an increasing number of web sites has started offering their data over the web in standard formats and exposed their functionality as APIs. A new type of applications has taken advantage of the new data and services available by mixing them, in order to generate new applications fast and efficiently, getting its name from its own architectural style: mashups. A set of applications that aims to help a user create, deploy and manage his mashups has also emerged, using various approaches. In this paper we discuss the key factors that should be taken into consideration when designing a mashup creator, along with the most important challenges that offer a field for research.**

## I. INTRODUCTION

Over the last decade there has been an explosion in the use of new web technologies that has lead to an increase of the available data and functionality over the Web. Numerous attempts have been made to utilize this increasing volume. Service-oriented architectures, portals and, recently, mashup technologies aim to create space for combinations of such functionality and data. Although the term mashup remains to this day loosely defined, in most cases it describes the act of combining functionality and data from different sources. In the past few years, there have been efforts to make such acts easily accessible to individuals with less technical expertise, safe, and, above all, available at lower cost than the traditional programming approach requires.

Quoting [1], "mashup makers (creators) are development environments for mashups". However, the notion of a mashup is used in many diverse ways. In general, mashups are applications, either standalone or embedded in an environment, that allow integration of different data sources and/or services that are available on the Web. The first mashups used to combine geographic data with online map services, but the combination of data/services has extended to different areas of interest. Mashup creators are tools that intend to facilitate the process of combining data and services, and – in some cases – offer the environment for executing the resulting mashup.

In this paper we inspect the key factors that should be taken into account when designing a mashup creation environment. We are, then, going to explore a range of attempts made to create such environments, focusing on the way the design aspects were considered. Finally, we discuss the challenges that are yet to be fulfilled.

## II. ASPECTS OF MASHUP CREATOR DESIGN

The solutions currently offered by mashup creation tools focus on subsets of the problem of mashup creation/generation, each addressing problem-specific needs. In this paragraph we attempt to define the problems and decisions that play a major role in designing a mashup creation tool.

### A. Mashup Lifecycle

One important aspect is the life duration of the resulting mashup. The life cycle of a mashup can span from minutes to years, depending on various factors. Short-lived mashups are used mostly in cases where the user wants to combine data from different sources in order to get a helpful visualization of the data. Such cases can include data sources available on the Internet (i.e. maps and data) or information stored in enterprise systems (databases, SAP etc), in order to generate quick reports. Long-lived mashups are usually part of rich Internet applications. Another factor that affects the duration of a mashup is the availability or changes to its building blocks. Services used to compose a mashup may stop functioning or be exposed using a different protocol, while the format that data are provided can change.

### B. Execution environment

Another major distinction is the actual execution environment of the mashup. The first mashups were deployed as web applications. The actual mash of data and/or services was performed using one of three alternate techniques:

- on the server side using the underlying software platform (i.e. J2EE, .NET, Rails),
- on the client side (the browser) by utilizing JavaScript and XMLHTTPRequest or
- a hybrid solution utilizing both of the previous solutions.

However, different execution environments have emerged. Although the browser still plays an important role, it is utilized in different ways. Resulting mashups can be applications based on frameworks such as Microsoft's Silverlight, JavaFX and Flash or can be browser-specific plug-ins that create a new local page eor customize an existing web page according to a script written in a scripting language (e.g. Greasemonkey, http://www.greasespot.net), in order to generate the mashup. Domain specific languages (mostly XML) have also appeared, offering the means to define the components of the mashup, their interactions and the operations required to generate the result. DSLs are commonly used as an intermediate format for describing a mashup, with an underlying platform performing the actual operations. The major advantage of mashups described in a DSL is that they are portable and reusable. Generated mashups can be stand-alone applications running on personal computers, mobile phones and other Internet-enabled devices, with some of the mashup creators allowing exporting the final mashup in more than one of the above forms.

Another approach is to allow mashups to be added to an existing site as applications or widgets, allowing the user to customize the appearance and functionality of his personal page. In such cases, the user's homepage becomes a dashboard that he can manage by adding the building blocks that are offered to him. Mashups added to his page can be either isolated or interact with each other, depending on the underlying architecture.

### C. Intended user audience

An important point of variation among mashup creators is the user level. Different mashup creators target different audiences and the experience they have in creating applications or mashups. In [1] the users are divided into three categories, depending on their programming skills. The first group contains users with good programming skills that can use a mashup creator, in order to prototype their mashup or save development time. The second group consists of users that understand the notion of data types and flows, while the third group represents the ones with average understanding of computers and Internet. Most mashup creators try to leverage off of the users the effort required to code a mashup. The most common approach is using a GUI design tool. However, this doesn't always simplify the mashup creation process for all the types of users equally. For example, a GUI design tool might require the user to be familiar with the concepts of regular expressions, XPath or other technologies. Alternatives have been proposed by various mashup creators, including the idea of mashup by example [6] (or sample [3]). In this case, the mashup creator asks the user to navigate the Web pages containing the data he will use and it attempts to collect information for available data sources or services.

### D. Data Source and Service Discovery

The key ingredient for building mashups is to combine data and services. In order for mashup creators to be efficient, they need to know which services and data the user wants to use. Different approaches include repositories, gathering information by allowing a user to browse the sites he wants to use or by providing technical details directly.

Repositories provide a catalogue of services and/or data sources. Apart from the actual data/service metadata, the stored information may contain details, such as ratings, comments, similarity indicators that can help the user locate useful services. Indexing can be performed through using simple solutions like keywords or a hierarchical structure, or more complex, using ontologies to manage the relationships among data. These repositories can be either local, used by few users, or centralized, offering advanced functionality for exposing and sharing services and mashups.

Programming by example is a different approach that – as mentioned above – targets mostly users with less programming knowledge. The user visits the Web pages containing the data. Simpler solutions allow defining the desired data by selecting them, while more intelligent ones attempt to understand the page by analyzing its structure or intercepting calls to remote services.

A feature offering added value to mashup creators is to enable reuse of previously created mashups. Existing mashups can be customized or used as building blocks for new ones, either by directly embedding them in the user interface or by invoking the functions that generate the results presented on it.

### E. Supported technologies

Content sources and services available on the Web can be accessed using a plethora of technologies. Data can be accessed in multiple forms, with XML based standards like Atom and RSS being extremely popular. However, it can also be retrieved in various formats (i.e. JSON, microformats) or directly as HTML by performing simple HTTP requests. Other resources include comma-separated files, spreadsheet programs or even databases. Services are also exposed using various standardized protocols like REST or SOAP. A mashup may be composed by any of these technologies, or even from another, non-standard, limited by the restrictions applied on the execution environment and the capability of the mashup creator to support these technologies. Another popular feature is to allow a mashup to obtain data from internal resources, such as Excel files or databases using SQL.

### F. Consumer-oriented or Enterprise

Another distinction can be made by the target group of the resulting mashups. The first mashup creators aimed to allow resource composition for resources available on the Web. However, the ability of mashups to combine data and generate information on demand, led to the design of mashup creators aiming for the enterprise sector. Enterprise mashups can also be seen as an extension of Service Oriented Architecture (SOA), allowing orchestration of both external (available on the Web) and internal (databases, enterprise services) resources, in order to offer situational applications that can help in decision-making. Another distinction between the two types of mashup creators is made in [2], mentioning that consumer-oriented aim mainly towards individuals that want to aggregate data i.e. for their homepage, while enterprise mashups target members of the enterprise that can access more stable resources.

## III. ASPECT INTERRELATIONS

Although the above aspects are important on their own, it is important to note that there's a connection among some of them. Design decisions on one of them may affect one or more of the rest.

One of the most important decisions, when designing a mashup creation tool, is if it will be consumer or enterprise oriented. Internal resources in enterprises offer more well-structured data, obey well-established standards (i.e. SOA related standards) and are more persistent than the external resources. Enterprise mashup creators should provide the means for utilizing these standards, while their persistent nature allows them to be indexed in repositories, allowing faster discovery. Repositories can, also, store the mashups, in order to enable reuse or fast access from other members of the enterprise. Reusability requirements and higher availability of the mashed resources usually result in longer life spans of the resulting mashup.

The lifecycle of the resulting mashup may interact with the execution environment. Apart from when and how the implementation requires data to support the existence of the service, if the repositories remain unavailable, the whole of the mashup will remain loose. This impact may lead to the decision which concludes in the election of server-side data execution, when we intend to produce long-term mashup applications.

The target user level is, also, one of the most affecting aspects. There's usually a tradeoff between the simplicity and the level of customizations that can be performed by the mashup creator. Targeting users with average understanding of computers limits the technologies involved in the mashup creation to the ones that the user can understand in a visual manner (i.e. extract data only from HTML), or the execution environment, while it leads to short-lived mashups that are not reusable.

## IV. MASHUP CREATORS

The elements described earlier in this paper have affected the design of the most popular mashup creation tools. Although development in some of them has terminated or is in pause, we focus on how design decisions have affected the result and not on the success of the result.

Yahoo Pipes! (http://pipes.yahoo.com) borrows the metaphor of piping, common in the unix world, indicating its functionality. It is a tool that allows you to create data mashups by combining and manipulating feeds, before outputing them in a manner, nearly as fast and as convenient as in the unix pipes. While these mashups can be used only once, the application allows the mechanism to be saved, published and re-used as a feed in new mashups. The application, also, creates space for situational uses of the feeds created, by providing the options to geocode the feed or gets the output in forms acceptable by map visualizations such as KML.

In another approach proposed by d.mix ([3]), the time spent on mashup data was considered the most important factor. To make the generation of mashups rapid, they employed a mechanism divided into two parts. The first part was the ability of the environment, utilizing a service-to-site map, to track the service calls made by a web site's elements by monitoring the elements selected visually (e.g. click) by the user. The second was a server-side active wiki, which hosted scripts that could be accessed via the browser and be personalized. This technique, called programming by sample, has aimed to allow faster mashup creation than the similar programming by example. Thus, d.mix focused on the ability to create short-lived mashups, while providing the ability to add code which could extend its functionality.

In a similar fashion, Intel Mash Maker ([4]) comes as a plug in to the browser and runs mostly on the client side. This enables it to be able to monitor user activity and, thus, to provide more personalized suggestions on how content can be mashed while, the user is browsing the Web. Mashups themselves are described internally by a functional language. The application utilizes a collaboratively maintained database of extractors, in order to assist the user in extracting data from the raw HTML of particular kinds of pages. The extractor database can be edited by the users, while on the other hand it keeps track of changes, so they won't disable previous functionality stored in the database.

Another effort to create a mashup environment was Vegemite, presented in [5]. It combined a somehow improved version of CoScripter plugin with a set of tables (VegeTables), providing a client side environment that could record user's actions, represent them as scripts in a language resembling English and re-run them on the set of data placed in the tables. The tables utilized an XPath scheme to extract rules and relations among the data put there by the user. While running mostly on the client side, Vegemite provided a mechanism to store scripts in a central repository, in order for them to be reused by other users. Similarly, it provided a mechanism for the table instances to be stored in a repository, in order to be available from different computers by the same user.

Karma ([6]), on the other hand, focused on automating the process of data extraction from Web sites. Relying heavily on XPath, it proposed a mechanism to extract the data, utilize a model on them, bring them in a consistent form, create rules so that they can be integrated with other data and, finally, proposed a method to visualize them.

Presto (available at http://www.jackbe.com/products/) is an enterprise mashup creation environment offered by JackBe. Its focus is mostly on enterprise mashups that can be built in a matter of "3 clicks versus 3 months". To achieve this, presto includes a sophisticated visual environment, allowing the combination of resources drawn from inside or outside the enterprise, which are presented as parts of the GUI. JackBe offers many service calls modeled as building blocks in the GUI, which the designer can configure using their various parameters. It, also, offers the ability to describe a service call in the visual environment and edit its output..

TABLE I. MASHUP CREATOR OVERVIEW

| Mashup Creator | Life span | Execution Environment | Data and service origin | User level | Range of supported technologies | Consumer / Enterprise |
|---|---|---|---|---|---|---|
| Yahoo Pipes! | Medium to long | Server-side execution[a] | User defined feeds | Simple to Intermediate | Limited | Both |
| d.mix | Short | Browser (as plugin) | By browsing HTML | Simple | Limited | Consumer |
| Intel Mash Maker | Medium | Browser (as plugin) | By browsing HTML | Simple | Limited | Consumer |
| Vegemite | Short to medium | Browser (as plugin) | By browsing HTML | Simple | Limited | Consumer |
| Karma | Short | Stand-alone application | By browsing HTML | Intermediate | Limited | Consumer |
| Presto | Medium to long | Server-side execution[b] | Repository, mashup enablers | Intermediate to advanced | Extensive | Enterprise |

a. On Yahoo's servers
b. Using Presto Enterprise Mashup Server

The main concept of the GUI is similar to Yahoo Pipes! wiring, but its functionality is far from being limited to feeds. It, also, offers the ability to extract data from sources outside the Web through the use of various "connectors" (http://www.jackbe.com/products/connectors.php). In the meantime, it has many ways of visualizing the data in the stages before the mashup is completed, thus it makes the design easier. Finally, it provides mechanisms to visualize the result of the mashup

## V. CHALLENGES

Mashups seem to have created a rush of interest, especially after the popularity enjoyed by Google Maps. This excitement was directed to certain areas that could enhance mashup making and, perhaps, create broader mashup developer communities. In order to turn this interest into advantage, a mashup environment should be designed carefully, considering various factors that are relevant to Web application design, but remain somewhat different as well. The mashup creator environment can be viewed both as a system, being an application, but also as something more. Its users constitute a network that not only uses the system, but can also expand its functionality and enrich its databases.

### A. Mashup Creator Environment as a System

From a designer's perspective, the mashup creator environment can be viewed as a system which has a number of acceptable input types, is able to perform a number of transformations on them and, finally, outputs a result. In this context, we attempt to present some of the challenges faced when designing a mashup creator environment.

*1) Input*

The first challenging part of designing such an environment is the way it treats its data input. The broader the variety of the data types the application is able to support, the more diverse types of mashups it can produce. Seeing data as the input in such a system brings forth three major difficulties. The first is that the data available on the Web can be in forms that are not predictable. This fact raises the difficulty of assessing them to produce a standard output. The second is that data can be in places that are and should remain protected, such as intranets or sites that require some kind of authentication to access them. Some solutions provided by the applications examined were either intelligent algorithms for data extraction on the first part of the problem or another level of abstraction that could treat data input, regardless of their type, leaving the specifics to other pieces of software. The third difficulty is the availability of data. There are many cases where the mashup creating software acts as the middle man between data and the mashup. This is where there must be a way to monitor changes in the sources, so that mashups remain functional.

The other aspect of mashup resources is services provided by different vendors in the form of APIs. This, also, presents challenges similar to those data retrieval, integration and sustainability does. Services can be exposed either in standardized ways or using custom protocols and technologies. It is, also, extremely possible (if not sure) that new standards will arise. On many occasions, instead of performing message exchange over an abstract protocol, the mashup creation requires embedding scripts written on a specific programming language (such as JavaScript for Google maps). It is also usual for services to require some kind of configuration (i.e. developer key), which the environment should somehow be able to handle. This is also something that raises the complexity and, perhaps, should be considered as a separate aspect of the design, maybe handled by a repository or some similar mechanism. Finally, similarly to the data, services are bound to change their availability status in various ways. The environment should be able to monitor these changes in status and provide solutions or alternatives to its users, both in terms of design and execution time.

*2) Output*

The way a mashup environment structures its output is important in various ways. Output can be in a format that can be consumed by other applications or it can be an application by itself. In the first case, the flow of information on the Web is facilitated, while in the second one more logic, in terms of quantity or diversity, is added to the Web.

Regardless of the above classification, the final form of the mashup plays an important role on whether the mashup can be reused as a building block by itself in the environment and, in this way, extending the environment's building blocks for new mashups or if it can be deployed or integrated in sites outside the environment.

As seen in many of the consumer mashups that have appeared up to this day, there are many limitations presented either by the client-server architectural models or by the mechanisms employed, in order to make the environment usable by people with little or no technical expertise. On the other hand, enterprise mashup creating environments have to face important security limitations when they attempt to combine internal to the enterprise data and functionality with those available on the Web.

Perhaps the most important challenge, in terms of the reusability of the mashups produced, either in a consumer or in enterprise environments, is the lack of standardized ways to achieve this production. This is considered to be one of the most important limitations of mashup creation by most of the vendors. Some of them, in order to overcome this limitation, have joined their forces towards creating an Open Mashup Alliance ([7]), which aims to the creation of an open standard for achieving mashup creation. Whether this effort is going to facilitate innovation or not is still under debate due to various issues, such as current licensing.

*3) Transformations*

The way a mashup creator functions upon its input is quite strictly related to the mechanisms it provides to its users to manipulate input through the user interface. UI can act as a limiting factor, especially due to the mashup making main concept: it should require less time than it would if a similar application were to be developed by a programmer. The effort to keep the way the environment's functionality is presented to the users under graspable concepts acts in an antagonistic manner to the type of the functionality itself.

This is a good reason why many of the mashup making environments (either consumer or enterprise) failed to gather people into big enough communities to support and extend them. On the other hand, through realizing this and focusing on providing solutions by functionality that aims to smaller problems can produce a good, yet limited, environment.

Again, this antagonistic relation between what concept (represented in the UI) is graspable by the users and the functionality this can support is amplified by the lack of standards that could loosen the relations between those two design aspects.

*B. Mashup Creator Environment as a Network*

Regarding the environment's users, the designer should take into account not only how it will be accessible and easy to use, but how it will also facilitate the exchange of information between the users in manners that can expand its functionality and spark innovation.

Between the various definitions of what a mashup really is exists a difficulty in defining the user base of an environment aiding the creation of mashups. During design time, the target group should be decided. This is, among other things, a decision about the kind of the community that will later be formed around the environment. Taking into account that technologies around mashup making are far from mature we assume that carefully selecting and attracting the right user base and managing to maintain an active community around the tool is vital not only for the tool's survival and growth, but also for the evolution of the technologies involved.

VI. SUMMARY

The present paper is a review of our research on an attempt to create our own mashup creator environment. While numerous attempts have been made in the past, we reference only a number of them as examples of the designing aspects, the difficulties involved and how they can be overcome. We believe that mashup making environments can attract research on various fields, since the technologies involved are still far from mature. Open Mashup Alliance is a first step to create an Open Standard and to form an active community around it. The evolution of the browsers, as envisioned by Google, will play a serious role in future mashup making environments. Finally we expect that many of the design difficulties will be overcome by the adoption of HTML 5.